\documentclass[reprint,superscriptaddress, a msmath,amssymb,aps,prb]{revtex4-2}

\usepackage{amssymb}
\usepackage{amsmath}
\usepackage{graphicx}
\usepackage{dcolumn}
\usepackage{bm}
\usepackage{color}
\usepackage{threeparttable}

\newcommand{\lBEST}{$\lambda$-BEDSe}
\newcommand{\lET}{$\lambda$-ET}
\newcommand{\lSTF}{$\lambda$-STF}
\newcommand{\lBETS}{$\lambda$-BETS}
\newcommand{\doublecirc}{{\ooalign{$\bigcirc$\crcr\hss$\circ$\hss}}}

\begin{document}

\preprint{} 

\title{Antiferromagnetic ordering of organic Mott insulator $\lambda$-(BEDSe-TTF)$_2$GaCl$_4$}

\author{A.~Ito}
 \affiliation{Graduate School of Science and Engineering, Saitama University, Saitama, 338-8570, Japan}
\author{T.~Kobayashi}
 \email{tkobayashi@phy.saitama-u.ac.jp}
 \affiliation{Graduate School of Science and Engineering, Saitama University, Saitama, 338-8570, Japan}
\author{D.~P.~Sari}
 \affiliation{Innovative Global Program, College of Engineering, Shibaura Institute of Technology, Saitama 337-8570, Japan}
 \affiliation{Advanced Meson Science Laboratory, RIKEN, Saitama 351-0198, Japan}
\author{I.~Watanabe}
 \affiliation{Advanced Meson Science Laboratory, RIKEN, Saitama 351-0198, Japan}
\author{Y.~Saito}
 \affiliation{Institute of Physics, Goethe-University Frankfurt, 60438 Frankfurt (M), Germany}
\author{A.~Kawamoto}
 \affiliation{Department of Condensed Matter Physics, Graduate School of Science, Hokkaido University, Sapporo 060-0810, Japan}
 \author{H.~Tsunakawa}
 \author{K.~Satoh}
  \altaffiliation{Deceased, 11 June 2021}
\author{H.~Taniguchi}
 \affiliation{Graduate School of Science and Engineering, Saitama University, Saitama, 338-8570, Japan} 

\date{\today}

\begin{abstract}
The band structure and magnetic properties of organic charge-transfer salt $\lambda$-(BEDSe-TTF)$_2$GaCl$_4$ (BEDSe-TTF: bis(ethylenediseleno)tetrathiafulvalene; abbreviated as $\lambda$-BEDSe) are investigated.
The reported crystal structure is confirmed using X-ray diffraction measurements, and the transfer integrals are calculated. 
The degree of electron correlation $U/W$ ($U$: on-site Coulomb repulsion, $W$: bandwidth) of $\lambda$-BEDSe is larger than one and comparable to that of the isostructural Mott insulator $\lambda$-(ET)$_2$GaCl$_4$ (ET: bis(ethylenedithio)tetrathiafulvalene, abbreviated as $\lambda$-ET), whereas the $U/W$ of the superconducting salt $\lambda$-(BETS)$_2$GaCl$_4$ (BETS: bis(ethylenedithio)tetraselenafulvalene) is smaller than one. 
$^{13}$C-NMR and $\mu$SR measurements revealed that $\lambda$-BEDSe undergoes an antiferromagnetic (AF) ordering below $T_{\rm N} = 22$~K. 
In the AF state, discrete $^{13}$C-NMR spectra with a remaining central peak are observed, indicating the commensurate AF spin structure also observed in $\lambda$-ET. 
The similarity between the structural and magnetic properties of $\lambda$-BEDSe and $\lambda$-ET suggests that both salts are in the same electronic phase, i.e., the physical properties of $\lambda$-BEDSe can be understood by the universal phase diagram of bandwidth-controlled $\lambda$-type organic conductors obtained by  donor molecule substitution. 
\end{abstract}

\maketitle

\section{Introduction}
\begin{figure}[tbp]
\begin{center}
\includegraphics[width=8cm]{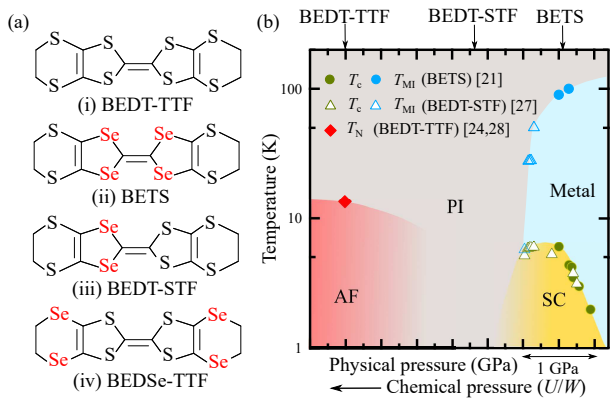}
\end{center}
\caption{
(a) Molecular structure of (${\bf i}$)BEDT-TTF, (${\bf ii}$)BETS, (${\bf iii}$)BEDT-STF, and (${\bf iv}$)BEDSe-TTF. 
(b) Phase diagram of $\lambda$-$D_2$GaCl$_4$ ($D$ = BEDT-TTF, BEDT-STF, BETS). 
$T_\mathrm{c}$ (green symbols) for SC transition and $T_\mathrm{MI}$ (blue symbols) for metal--insulator transition were determined from the resistivity measurements under pressure \cite{Tanaka1999,Minamidate2015}. 
The AF ordering temperature $T_\mathrm{N}$ (red rhombus) of \lET\ was obtained by a $^{13}$C-NMR and electron-spin resonance measurements \cite{Mori2001,Saito2018a}.  
}
\label{fig1}
\end{figure}

In molecular-based organic conductors, electronic properties are drastically changed by applying physical pressure. 
For example, (TMTSF)$_2$PF$_6$ (TMTSF: tetramethyltetraselenafulvalene) is a quasi-one-dimensional metal exhibiting a spin-density-wave ordering below $12$~K \cite{Andrieux1981}, 
and $\kappa$-(BEDT-TTF)$_2$Cu[N(CN)$_2$]Cl [BEDT-TTF (ET): bis(ethylenedithio)tetrathiafulvalene (Fig.~\ref{fig1}(a)-${\bf i}$)] is a quasi-two-dimensional Mott insulator exhibiting an antiferromagnetic (AF) ordering below $22.8$~K \cite{Kawamoto1995,Miyagawa1995,Ito2015}. 
However, they both show superconductivity under pressure \cite{Jerome1980,Williams1990}.
Therefore, it is believed that electron correlation plays a key role in the occurrence of superconductivity, and to this end, these systems have been well studied \cite{Brown2015}.
To understand the relationship between superconducting (SC) and adjacent electronic phases, each phase should be investigated in detail, and the pressure--temperature phase diagram should be established.
For this purpose, chemical pressure can be applied using molecular substitution.
Replacing anion molecules PF$_6$ with ClO$_4$ and Cu[N(CN)$_2$]Cl with Cu[N(CN)$_2$]Br or Cu(NCS)$_2$ results in a pressure effect, which leads to superconductivity under ambient pressure \cite{Bechgaard1981,Urayama1988,Kini1990}.
This approach greatly facilitates the understanding of superconductivity. Thus far, many unconventional properties have been reported \cite{Lang2004}.

$\lambda$-(BEDT-TSF)$_2M$Cl$_4$ [BEDT-TSF (BETS): bis(ethylenedithio)tetraselenafulvalene (Fig.~\ref{fig1}(a)-${\bf ii}$), $M$ = Ga, Fe] also exhibits interesting properties such as field-induced superconductivity \cite{Uji2001}, Fulde--Ferrell--Larkin--Ovchinnikov superconductivity \cite{Tanatar2002b,Coniglio2011,Uji2015}, and an anisotropic SC gap \cite{Balicas2001,Imajo2016,Imajo2019,Kobayashi2020a}. 
Thus, they should be investigated in addition to the above-mentioned compounds.
To investigate the mechanism of superconductivity in $\lambda$-type salts, the chemical pressure effect of substituting bromine for chlorine in $\lambda$-(BETS)$_2$GaCl$_4$ (abbreviated as \lBETS) has been studied \cite{Kobayashi1997,Tanaka1999}. 
With an increase in the amount of bromine, the spin-density-wave phase has been found to be adjacent to the SC phase \cite{Kobayashi2020}. 
However, in the system of $\lambda$-(BETS)$_2$GaBr$_x$Cl$_{4-x}$, the pressure range investigated by bromine substitution is narrow because $\lambda$-type salts can be obtained only in the range $x < 2$ \cite{Tanaka1999} and superconductivity occurs at $0.12$~GPa for $x$ = $1.5$ \cite{Tanaka1996}.

For complementary information on a wider pressure range, a universal phase diagram using donor molecular substitution has been proposed, as illustrated in Fig.~\ref{fig1}(b) \cite{Mori2001,Phasediagram}.
As demonstrated by the substitution of TMTSF for TMTTF (tetramethyltetrathiafulvalene) \cite{JEROME1991}, the substitution of S and Se in the TTF skeleton leads to a large pressure effect because of the significant change in the intermolecular transfer integrals. 
In fact, $\lambda$-(BEDT-STF)$_2$GaCl$_4$ [BEDT-STF (STF): unsymmetrical-bis(ethylenedithio)diselenadithiafulvalene (Fig.\ref{fig1}(a)-${\bf iii}$); abbreviated as \lSTF], is an insulator at ambient pressure, and superconductivity emerges at a pressure of $\sim 1.3$~GPa \cite{Mori2001,Minamidate2015}, which confirms the universal phase diagram.
$\lambda$-(ET)$_2$GaCl$_4$ (abbreviated as \lET), which is located at a more negative pressure side than \lSTF, undergoes a transition from a Mott insulator to an antiferromagnet at $13$~K \cite{Saito2018a}, whereas magnetic ordering is not observed in \lSTF\ \cite{Minamidate2015,Saito2019}.
The absence of the magnetic order may be the quantum critical effect of \lSTF\ being located between the AF and SC phases \cite{Saito2019} or the disorder effect originating from the asymmetric BEDT-STF molecules.
In contrast, the AF order has been observed in $\lambda$-(BETS)$_2$FeCl$_4$ \cite{Akutsu1997,Oshima2017} and $\lambda$-(STF)$_2$FeCl$_4$ \cite{Minamidate2018,Fukuoka2018,Fukuoka2020}, although there is a contribution of $3d$ spins from the Fe ions. 
Physical properties in \lET\ that do not contain asymmetry in its donor molecule should be investigated under pressure to understand why \lSTF\ does not exhibit any magnetic ordering.
However, several polymorphs are obtained simultaneously in the synthesis of \lET, and the main product is $\delta$-(ET)$_2$GaCl$_4$ \cite{Zorina2001,Zhang2013,Kurmoo1996}, which complicates the study of the physical properties of \lET. 
Further, several polymorphs of (ET)$_2$FeCl$_4$ have been synthesized \cite{Mallah1990,Zhang2010}, but the $\lambda$-type salt is not obtained.

To address the aforementioned problems, we focused on the BEDSe-TTF molecule shown in Fig.~\ref{fig1}(a)-${\bf iv}$, where BEDSe-TTF denotes bis(ethylenediseleno)tetrathiafulvalene.
Cui \textit{et al}. reported the lattice parameters of $\lambda$-(BEDSe-TTF)$_2$GaCl$_4$ (\lBEST) \cite{Cui2005a}, which are larger than those of \lET, indicating a negative chemical pressure effect. 
The negative chemical pressure effect by BEDSe-TTF molecular substitution for an ET molecule was demonstrated in $\kappa$-(ET)$_2$Cu[N(CN)$_2$]Br \cite{Sakata1998}. 
These results suggest that \lBEST\ is a promising candidate to provide further information on the insulating phase of the universal phase diagram.
Moreover, there are no reports of polymorphisms in either (BEDSe-TTF)$_2$GaCl$_4$ or (BEDSe-TTF)$_2$FeCl$_4$, which is a great advantage in investigating their physical properties.
However, physical properties are yet to be reported for \lBEST\ except for the semiconducting resistivity above $200$~K \cite{Cui2005a}. 

In this study, we investigate the structural and magnetic properties of \lBEST\ to explore whether it is located on the universal phase diagram and to promote the understanding of the insulating phase.  

\section{Experimental}
\label{exp}
Single crystals of \lBEST\ were prepared by the electrochemical oxidation of BEDSe-TTF in a solution of chlorobenzene containing $10$~\% ethanol with tetrabutylammonium salt of GaCl$_4^{-}$.
The samples were needle-like crystals as in the other $\lambda$-type salts. 
Polymorphism was not confirmed in our experiments unlike for other $\lambda$-type salts. 

Single-crystal X-ray diffraction measurements on $\lambda$-$D_2$GaCl$_4$ ($D$ = BEDSe-TTF, ET, and BETS) were performed using a Bruker SMART APEX2 diffractometer by employing a graphite-monochromated Mo-$K_{\alpha}$ radiation ($\lambda$ = $0.71073$~\AA) at the Comprehensive Analysis Center for Science, Saitama University, Japan.
The diffraction data were collected at $110$~K, and the structures were solved using SHELXT \cite{Sheldrick2015a} and refined using SHELXL \cite{Sheldrick2015}.

The overlap integrals, band dispersion, and Fermi surface were obtained by the tight-binding calculation based on the extended H\"{u}ckel method \cite{Mori1984}.
The transfer integrals $t$ were estimated from the overlap integrals $S$ assuming that $t$ = $ES$, where $E$ represents a constant of $-10.0$~eV.
For Se-containing organic conductors, the choice of the H\"{u}ckel parameters for the Se atom remain controversial \cite{Whangbo1982,Grant1982}. 
Mori and Katsuhara studied the parameter dependence of the overlap integrals in $\lambda$-type salts, and in this study, we applied the same parameter set \cite{Mori2002,Huckelparameter}. 

Magnetization measurements on \lBEST\ were performed using a superconducting quantum interference device magnetometer (Quantum Design MPMS XL-7).
The magnetic susceptibility of polycrystalline samples with a weight of $7.9$~mg was measured under a magnetic field of $1$~T between $2$ and $300$~K. 
The spin susceptibility was acquired by subtracting the core diamagnetic contribution of $-5.00 \times 10^{-4}$~emu/mol estimated from the measured susceptibility of ingredients such as neutral molecules.

Muon-spin rotation ($\mu$SR) experiments on \lBEST\ were carried out using a general purpose surface-muon instrument at Swiss Muon Source (S$\mu$S), Paul Scherrer Institut (Villigen, Switzerland). We used a continuous muon beam with the spin polarization parallel to the beamline.
A randomly oriented polycrystalline sample of $50$~mg was wrapped in silver foil. 
Measurements were conducted under zero magnetic field at temperatures between $40$--$1.6$~K to cover the temperature range of the magnetic transition.

\begin{figure}[tbp]
\begin{center}
\includegraphics[width=8cm]{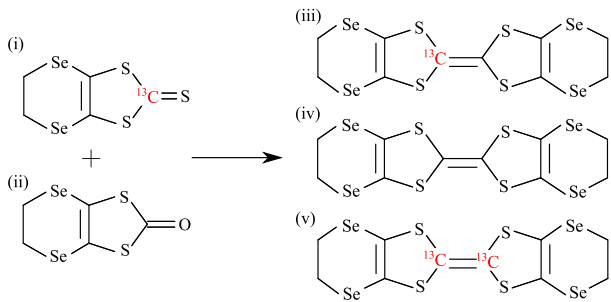}
\end{center}
\caption{Synthesis of $^{13}$C enriched BEDSe-TTF molecules.}
\label{fig2}
\end{figure}

For a $^{13}$C NMR experiment, we prepared $^{13}$C enriched BEDSe-TTF molecules synthesized from $^{13}$C enriched thioketone (${\bf i}$) and cool ketone (${\bf ii}$) (Fig.~\ref{fig2}) \cite{Wang1989}, as used in the preparation of the single-site $^{13}$C-enriched ET molecule \cite{Kawamoto2004,Matsumoto2012}. 
This cross coupling afforded $78$~\% of (${\bf iii}$), $18$~\% of (${\bf iv}$), and $4$~\% of (${\bf v}$). 
Their ratio was estimated from the cross-coupling reaction of deuterated thioketone (${\bf i}$) and cool ketone (${\bf ii}$) using mass spectroscopy.
As molecule (${\bf iv}$) is NMR inactive, the NMR signals were obtained from the molecule (${\bf iii}$), which helped prevent the Pake doublet problem \cite{Pake1948}. 
$^{13}$C-NMR experiments were performed for a single crystal with dimensions of $10 \times 0.35 \times 0.04$~mm$^3$ in a magnetic field of $6$~T parallel to the long axis of the BEDSe-TTF molecules, where the NMR shift becomes minimum in the $a^{*}b^{*}$ plane. 
The NMR spectra were obtained by the fast Fourier transformation of the spin-echo signals with a $\pi/2$-$\pi$ pulse sequence. 
The typical $\pi/2$ pulse length was $2$~$\mu$s.
Spin-lattice relaxation time $T_1$ was measured by a conventional saturation-recovery method.

\begin{figure*}[tbp]
\begin{center}
\vspace{1cm}
\includegraphics[width=2\columnwidth]{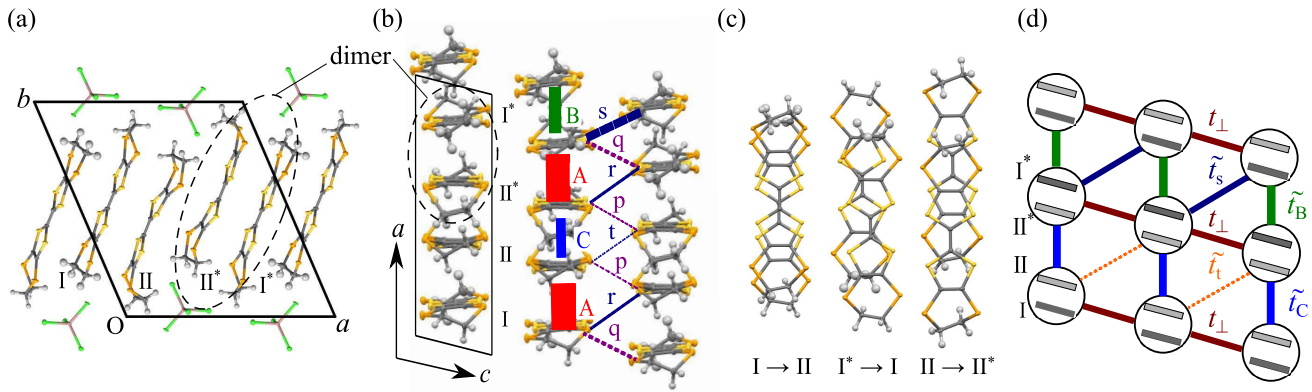}
\end{center}
\caption{
(a) Layered crystal structure of \lBEST~viewed along the $c$ axis.
(b) In-plane structure of the BEDSe-TTF layer in the $ac$ plane. 
Square and dotted ellipse represent the unit cell and dimer of the BEDSe-TTF molecules, respectively.
(c) Overlaps between each molecule viewed perpendicular to the molecular plane.
(d) Schematic representation of the BEDSe-TTF layer in a dimer model, where circles represent dimers.
}
\label{fig3a}
\end{figure*}
\section{Results and Discussion}
\subsection{Crystal structure}
\label{crystalstructure}
\begin{table}[tbp]
\caption{
Lattice parameters for $\lambda$-$D_2$GaCl$_4$ ($D$ = BEDSe-TTF, ET, and BETS).
}
\begin{threeparttable}
\begin{ruledtabular}
\begin{tabular}{cccc}
Parameter & \lBEST & \lET & \lBETS \\ 
\hline 
$a$ (\AA) & $16.1260(14)$ & $15.9661(15)$ & $15.9166(19)$ \\
$b$ (\AA) & $18.1099(15)$ & $17.9068(17)$ & $18.451(2)$\\
$c$ (\AA) & $6.6519(6)$ & $6.4544(6)$ & $6.5435(8)$\\
$\alpha$ ($^{\circ}$) & $97.397(1)$ & $98.520(1)$ & $98.606(1)$\\
$\beta$ ($^{\circ}$) & $97.154(1)$ & $96.731(2)$ & $95.960(1)$\\
$\gamma$ ($^{\circ}$) & $111.886(1)$ & $112.027(2)$ & $112.261(1)$\\
$V$ (\AA$^{3}$) & $1755.9(3)$ & $1661.4(3)$ & $1731.2(4)$\\
\end{tabular}
\end{ruledtabular}
\end{threeparttable}
\label{table1}
\end{table}

The structural analyses of $\lambda$-$D_2$GaCl$_4$ ($D$ = BEDSe-TTF, ET, and BETS) were performed at $110$~K because crystallographic data including atomic parameters were not reported in previous papers \cite{Mori2001,Cui2005a}. 
This information is useful not only for comparing the structure but also for performing band structure calculations. 
Table~\ref{table1} shows the lattice parameters of the three salts. Results are consistent with those previously reported \cite{Cui2005a,Tanaka1999,Mori2001}. 
These values show that the three salts are isostructural, which indicates that physical properties can be understood by the same phase diagram. 
From a comparison of the three salts, the unit cell volumes of \lBETS\ and \lBEST\ are larger than that of \lET\ by $4.2$~\% and $5.7$~\%, respectively. This implies that the replacement of S atoms with Se atoms in the ET molecule leads to lattice expansion.
In addition, the unit cell volume of \lBEST\ is larger than that of \lBETS, which shows that the substitution at the outer chalcogen atoms has a greater effect on the lattice expansion. 
These lattice expansions are considered the negative chemical pressure effect; however, \lBETS\ and \lBEST\ are metallic and insulating, respectively. 
To discuss the actual pressure effect, not only the lattice constants but also the intermolecular overlap integrals should be investigated. 

Figure~\ref{fig3a}(a)--(c) show the crystal structure of \lBEST. 
In this system, BEDSe-TTF layers and GaCl$_4$ layers are alternately stacked along the $b$ axis [Fig.~\ref{fig3a}(a)]. 
In the BEDSe-TTF layers, there are two crystallographically independent molecules: ${\rm I}$ (${\rm I}^*$) and ${\rm II}$ (${\rm II}^*$), where molecules marked with asterisks are related to the unmarked ones by the inversion center. 
These molecules are stacked along the $a$ axis [Fig.~\ref{fig3a}(b)]; however, the overlap modes characterized by sliding distance along the long axis of the molecule [Fig.~\ref{fig3a}(c)] and interplanar distance between the molecules are different. 
Table~\ref{tableX} shows that these values between molecules I and II are smaller than the others and are close to that of $\kappa$-type salts (sliding distance is $1.59$~\AA~and interplanar distance is $3.56$~\AA) \cite{Mori1999}, which suggests molecules I and II form a dimer. 

\begin{table}[tbp]
\caption{
Interplanar and sliding distances in the overlap mode between molecules for $\lambda$-$D_2$GaCl$_4$ ($D$ = BEDSe-TTF, BEDT-TTF, and BETS).
}
\begin{ruledtabular}
\begin{tabular}{cccc}
Modes & \lBEST & \lET & \lBETS  \\ 
\hline 
Interplanar distance & & &\\
I--II & $3.57$ & $3.49$ & $3.83$ \\
I--I$^*$ & $4.00$ & $3.99$ & $4.15$ \\
II--II$^*$ & $4.51$ & $4.11$ & $4.02$ \\
Sliding distance & & &\\
I--II & $1.00$ & $0.99$ & $1.20$\\
I--I$^*$ & $2.82$ & $2.68$ & $2.84$\\
II--II$^*$ & $4.80$ & $4.71$ & $4.76$
\end{tabular}
\end{ruledtabular}
\label{tableX}
\end{table}

Among the three salts, interplanar distances are significantly different whereas the sliding distances are insignificant (Table~\ref{tableX}).
These differences affect the magnetic interaction between dimers, as discussed in Sec.~\ref{Spinsus}.

\subsection{Band structure calculation}
We performed tight binding calculations for $\lambda$-$D_2$GaCl$_4$ ($D$ = BEDSe-TTF, ET, and BETS) using the obtained atomic parameters. 
Table~\ref{table2} shows the transfer integrals of the three salts, and the definitions are displayed in Fig.~\ref{fig3a}(b). 
As inferred from the overlap modes of the crystal structure, $t_{\rm A}$ is significantly larger than $t_{\rm B}$ and $t_{\rm C}$, which indicates that molecules I and II form a dimer from the perspective of electronic structure. 

The magnitude of the transfer integrals along the stack direction for \lBEST\ is intermediate between those for \lET\ and \lBETS.
In contrast, the transfer integrals perpendicular to the stack directions are not significantly different between \lET\ and \lBEST, and they are smaller than those for \lBETS. 
In the BEDT-TTF molecule, the electron densities of the inner chalcogen atoms are greater than those of the outer ones.
Considering that the chalcogen atom in the TTF skeleton of \lBEST\ and \lET\ is sulfur and that of \lBETS\ is selenium, 
transfer integrals perpendicular to the stack directions are dominated by the orbital overlap of inner chalcogen atoms, 
whereas the outer chalcogen atoms contribute to the transfer integrals along the stack direction.

\begin{table}[tbp]
\caption{
Transfer integrals $t$~($\times 10^{-3}$~eV) [Fig.~\ref{fig3a}(b)] of $\lambda$-$D_2$GaCl$_4$ ($D$ = BEDSe-TTF, BEDT-TTF, and BETS).
}
\begin{ruledtabular}
\begin{tabular}{cccc}
Parameter & \lBEST & \lET & \lBETS  \\ 
\hline 
$t_\mathrm{A}$ & $304.8$ & $282.1$ & $360.4$ \\
$t_\mathrm{B}$ & $-122.7$ & $-83.92$ & $-179.2$ \\
$t_\mathrm{C}$ & $-130.9$ & $-90.63$ & $-156.0$ \\
$t_\mathrm{p}$ & $20.04$ & $22.01$ & $30.51$ \\
$t_\mathrm{q}$ & $55.50$ & $46.63$ & $99.21$ \\
$t_\mathrm{r}$ & $55.68$ & $62.35$ & $123.7$ \\
$t_\mathrm{s}$ & $-121.3$ & $-132.9$ & $-173.2$ \\
$t_\mathrm{t}$ & $-24.84$ & $-16.56$ & $-25.78$ \\
\end{tabular}
\end{ruledtabular}
\label{table2}
\end{table}

\begin{table}[tbp]
\caption{
Upper part shows $U$ and $W$ ($\times 10^{-3}$~eV), and the ratio of $U$ to $W$ for $\lambda$-$D_2$GaCl$_4$ ($D$ = BEDSe-TTF, BEDT-TTF, and BETS).
The lower part shows the ratio of the transfer integrals of the dimer model $\tilde{t}_\nu$ ($\nu$: B, C, s, and t) to that of $t_{\perp}$~[Fig.~\ref{fig3a}(d)].
}
\begin{ruledtabular}
\begin{tabular}{cccc}
Parameter & \lBEST & \lET & \lBETS \\ 
\hline 
$U$ & $609.6$ & $564.2$ & $720.8$\\
$W$ & $500.4$ & $449.0$ & $809.1$\\
$U/W$ & $1.218$ & $1.256$ & $0.891$\\
 & & &\\
$|\tilde{t}_{\rm B}/t_{\perp}|$ & $0.935$ & $0.641$ & $0.707$  \\
$|\tilde{t}_{\rm C}/t_{\perp}|$ & $0.998$ & $0.692$ & $0.616$  \\
$|\tilde{t}_{\rm s}/t_{\perp}|$ & $0.924$ & $1.015$ & $0.683$  \\
$|\tilde{t}_{\rm t}/t_{\perp}|$ & $0.189$ & $0.126$ & $0.102$  \\
\end{tabular}
\end{ruledtabular}
\label{table3}
\end{table}

\begin{figure}[tbp]
\begin{center}
\includegraphics[width=8cm]{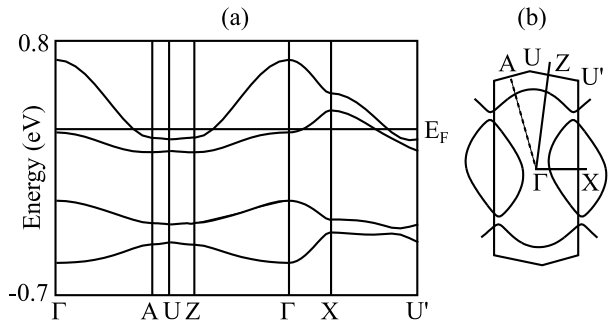}
\end{center}
\caption{
(a) Band structure and (b) Fermi surface of $\lambda$-(BEDSe-TTF)$_2$GaCl$_4$.
}
\label{fig3}
\end{figure}

From these transfer integrals, the band dispersion and Fermi surface of \lBEST\ were obtained as shown in Fig.~\ref{fig3}.
The band dispersion is split into the upper and lower bands as in the case of \lET\ and \lBETS\ because of the dimerized structure \cite{Mori2001,Saito2018a,Tanaka1999}.
The overlapped single Fermi surface is disconnected because of the anisotropic transfer integral lattice. 
The Fermi surface  consists of a two-dimensional cylindrical part and a one-dimensional flat part. These features are the same among all three salts.

Based on the discussion proposed by Hotta~\cite{Hotta2003}, we estimate the electron correlation by considering the transfer integrals of $\lambda$-type salts in the dimer model.
The transfer integrals in the dimer model are defined as  $\tilde{t}_{\rm B} \equiv t_{\rm B}/2$, $\tilde{t}_{\rm C} \equiv t_{\rm C}/2$, $\tilde{t}_{\rm s} \equiv t_{\rm s}/2$, $\tilde{t}_{\rm t} \equiv t_{\rm t}/2$, and $t_{\perp}\equiv (t_{\rm p} + t_{\rm q} + t_{\rm r})/2$ [see Fig.~\ref{fig3a}(d)].
Here, $t_{\perp}$ should be calculated carefully because the sign of the transfer integral between the dimers must be considered properly. 
In our calculation, the phase of the highest occupied molecular orbital is taken so that the intra-dimer overlap is negative, resulting in $t_{\rm A} > 0$.
As the hole resides on the antibonding molecular orbital in the dimer, the phase factor between the inter-dimer orbital corresponds to that used in the calculation; i.e., $t_{\perp} = (t_{\rm p} + t_{\rm q} + t_{\rm r})/2$.
Note that when the hole is on the bonding orbital, the phase of one of the molecules must be reversed, resulting in $t_{\perp} = (t_{\rm p} + t_{\rm q} - t_{\rm r})/2$. 

The on-site Coulomb repulsion energy is approximately proportional to the transfer integral within the dimer in a dimeric structure, i.e., $U = 2t_{\rm A}$ \cite{Kanoda1997c,Yoneyama1999}. 
The bandwidth $W$ is estimated according to the relation \cite{Hotta2003} 
\begin{equation}
\label{eq0}
 W = \sum_{\nu} \tilde{t}_{\nu} + 4t_{\perp} + \frac{(\sum_{\nu} \tilde{t}_{\nu})^2}{16t_{\perp}},
\end{equation}
where $\nu$ represents B, C, s, and t. 
Further, the $U$, $W$, and $U/W$ parameters of each salt calculated by these definitions are listed in the top part of Table~\ref{table3}. 
The three salts are situated at $U/W \sim 1$, indicating that they are in a region where itinerancy and localization are in competition. 
The $U/W$ parameters of \lBEST\ and \lET\ are nearly the same and greater than $1$, and that of \lBETS\ is significantly less than $1$.
These results suggest that \lBEST\ and \lET\ are more localized than \lBETS, and that there is a Mott transition between the two salts and \lBETS. These observations are consistent with experimental facts about the conductivity of $\lambda$-type salts \cite{Mori2001}. 

\subsection{Spin susceptibility}
\label{Spinsus}
\begin{figure}[tbp]
\begin{center}
\includegraphics[width=8cm]{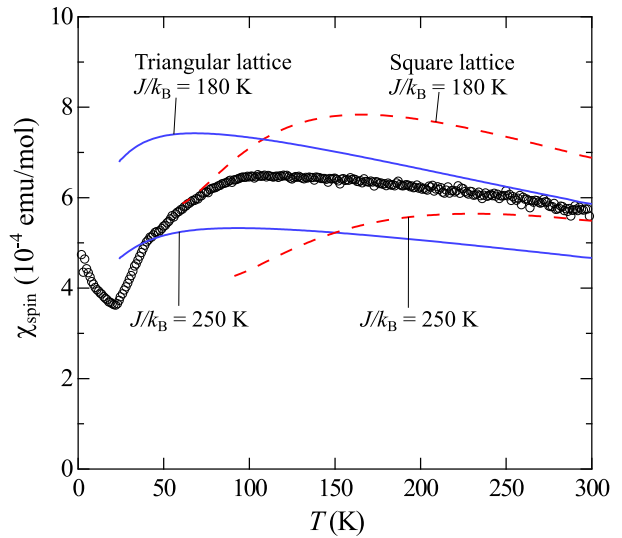}
\end{center}
\caption{
(a) Temperature dependence of spin susceptibility of \lBEST. 
Blue solid lines represent the 2D Heisenberg AF triangular-lattice model \cite{Tamura2002}, and the red dashed lines represent the 2D Heisenberg AF square-lattice model \cite{lines1970}. 
} 
\label{fig4}
\end{figure}
Figure~\ref{fig4} shows the temperature dependence of spin susceptibility $\chi_{\rm spin}$ of \lBEST. 
As the temperature is reduced from $300$~K, $\chi_{\rm spin}$ increases towards $100$~K, where it shows a broad maximum. 
Further decreasing temperature, $\chi_{\rm spin}$ decreases to $22$~K.
The broad maximum of $\chi_{\rm spin}$ is a characteristic of a system possessing a low-dimensional magnetic interaction network. 

For \lET~and \lSTF, the temperature dependence of $\chi_{\rm spin}$ have been discussed using the two-dimensional (2D) Heisenberg AF spin model \cite{Tamura2002,lines1970}.
Interestingly, the $\chi_{\rm spin}$ of \lET\ and \lSTF\ have been explained by the square and triangular lattice models, respectively \cite{Saito2018a,Saito2019,Minamidate2015}.
Here, we applied these analyses to \lBEST.
The solid and dashed lines in Fig.~\ref{fig4} show the temperature dependencies of $\chi_{\rm spin}$ assuming the triangular and square lattice AF spin models, respectively. 
As a rough estimation of the exchange interaction $J$, the calculation results for $J/k_B$ = $180$~K and $250$~K are shown, which are the upper and lower limits; here, the experimental data above $\sim 50$~K are included. Both models do not reproduce the experimental result.

Considering the network of $J$ between the dimers in \lBEST, we can speculate why the experimental result cannot be explained by these models.
Because $J$ is expressed as $J=4t^2/U$ in the case of localized systems, we discuss the network of $J$ using the transfer integrals relative to $|t_{\perp}|$, as shown in Table~\ref{table3}. 
The transfer integrals other than $|\tilde{t}_{\rm t}|$ are comparable to $|t_{\perp}|$, and $|\tilde{t}_{\rm t}|$ is negligibly small.
Thus, the network of the transfer integrals in \lBEST\ is a combination of a triangular ladder and a squared ladder, which is similar to the so-called trellis lattice discussed in Ref.~\cite{Sakakida2017} [see Fig.~\ref{fig3a}(d)].
This result is consistent with the result of spin susceptibility because the temperature of the broad peak of $\chi_{\rm spin}$ is roughly intermediate between those of the triangular and square lattice AF spin models.
Further calculations are required to verify whether the spin model of the  lattice can explain the experimental results.

Although \lET~and \lSTF~have almost the same structure as \lBEST, the simple square and triangular lattice AF spin models can reasonably explain the temperature dependence of the $\chi_{\rm spin}$ of \lET~and \lSTF. 
A possible reason for this is the difference in the network of $J$ between the three salts. There is actually a difference in the overlap modes in \lBEST, \lET, and \lBETS\ (Table~\ref{tableX}). 
$|\tilde{t}_{\rm B}/t_{\perp}|$ and $|\tilde{t}_{\rm C}/t_{\perp}|$ for \lET\ are smaller than those for \lBEST.
When $|\tilde{t}_{\rm s}|^2$ and $|t_{\perp}|^2$, being twice as large as $|\tilde{t}_{\rm B}|^2$ and $|\tilde{t}_{\rm C}|^2$, are dominant in the network of $J$, we can approximate \lET~as a square lattice, which is consistent with the experimental results of $\chi_{\rm spin}$ \cite{Saito2018a}.
For \lSTF, the transfer integrals are difficult to evaluate because of the molecular asymmetry of BEDT-STF. However, the difference in the network of $J$ between \lET~and \lBETS\ suggests that the network of $J$ in \lSTF\ is also different. 
Further, because line broadening of the $^{13}$C-NMR spectra has been observed in \lSTF\ \cite{Saito2019}, there is a possibility of charge disproportionation, which can modify the transfer integrals between the molecules.
These features may be responsible for the temperature dependence of the triangular lattice-like $\chi_{\rm spin}$ in \lSTF.

In this study, we systematically evaluated the magnitude of the relative transfer integrals in the dimer model to investigate the network of $J$.
Note that the relative transfer integrals depend on the calculation method and the H\"{u}ckel parameters \cite{Seo1997,Aizawa2018,Dita2021}. 
At least, a small $\tilde{t}_{\rm t}$ is a characteristic behavior of $\lambda$-type salts, and the lattice realized by neglecting $\tilde{t}_{\rm t}$ is considered to be the fundamental model for discussing the spin structure of $\lambda$-type salts. 
In this context, \lBEST\ would be a useful reference material for discussing the magnetism of $\lambda$-type salts. 

Below $22$~K, $\chi_{\rm spin}$ increases, while the magnitude of the increase is not as drastic as that observed in canted antiferromagnet $\kappa$-(ET)$_2$Cu[N(CN)$_2$]Cl \cite{Kawamoto1995,Ishikawa2018}. 
This can be attributed to the small amounts of magnetic impurities and/or magnetic transition where an anisotropy of magnetic susceptibility appears.
To clarify the magnetic state at low temperatures, microscopic measurements should be conducted.
As another anomaly, a small kink structure was observed at approximately $40$~K. This is a small change in slope and is observed in data extracted from preliminary magnetic torque measurements. Thus, although it may be intrinsic, its origin is unknown at this stage.

\begin{figure*}[tbp]
\begin{center}
\vspace{1cm}
\includegraphics[width=2\columnwidth]{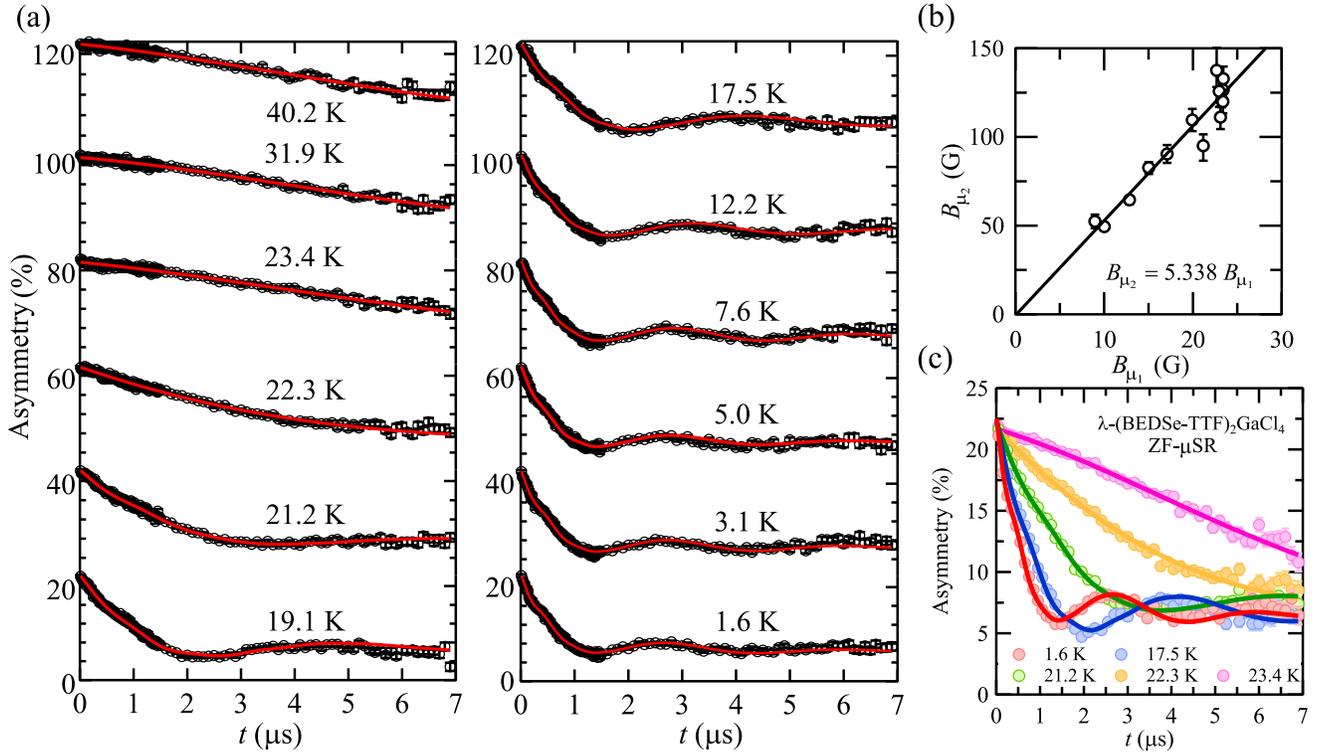}
\end{center}
\caption{
(a) Temperature evolution of $\mu$SR time spectra at zero magnetic field. 
Each spectrum is vertically shifted by $20$~\% for ease of comparison. 
Solid curves represent the fitting results obtained using Eq.~(\ref{eq1}) ($T \geq 23.4$~K) and Eq.~(\ref{eq2}) ($T \leq 22.3$~K).
(b) Relationship between $B_{\mu_1}$ and $B_{\mu_2}$ with the linear fitting. 
(c) Representative $\mu$SR time spectra. 
Solid lines are the fitting results when $B_{\mu_2} = 5.338B_{\mu_1}$ in Eq.~(\ref{eq2}).
}
\label{fig5}
\end{figure*}

\subsection{$\mu$SR}
A $\mu$SR measurement can be used for the sensitive detection of magnetic ordering to probe the magnetic state of \lBEST\ microscopically at low temperatures. 
Figure~\ref{fig5}(a) shows the time evolution of the muon-spin polarization ($\mu$SR time spectra) at several temperatures under a zero magnetic field. 
The $\mu$SR time spectra remain unchanged above $23.4$~K, below which the relaxation rate becomes larger and clear precession signals are observed, confirming a magnetic ordering.
To understand the observed $\mu$SR time spectra in detail, we analyze them separately for the paramagnetic and ordered phases as follows.

In the paramagnetic state above $23.4$~K, the $\mu$SR time spectra can be fitted by 
\begin{equation}
\label{eq1}
  A(t) = A \mathrm{e}^{-\lambda t}G_{\mathrm{KT}}(t)+A_{\mathrm{bg}},
\end{equation}
where $A$ and $A_{{\rm bg}}$ represent the relative ratios of amounts of the muons stopped inside the sample and in the silver sample holder, respectively, and $\lambda$ is the relaxation rate. 
$G_{\mathrm{KT}}(t)$ represents the Kubo--Toyabe function expressed as 
\begin{equation}
  G_{\mathrm{KT}}(t) = \frac{1}{3}+\frac{2}{3}\left( 1-\Delta^2 t^2 \right) \exp(-\frac{1}{2}\Delta^2 t^2 ),
\end{equation}
where $\Delta$ represents the distribution width of the depolarization rate of the nuclear spin contribution. 

In the ordered state, the main precession signals with a period of approximately $2$~$\mu$s and a kink at around $0.5$~$\mu$s were observed, although the latter is of small amplitude. 
To explain these $\mu$SR time spectra, we fitted them using the function
\begin{multline}
\label{eq2}
  A(t)=A_0 \mathrm{e}^{-\lambda_0t} + A_1\cos(\gamma_{\mu} B_{\mu_1} t+ \phi)\mathrm{e}^{-\lambda_1t} \\ 
  +A_2\cos(\gamma_{\mu} B_{\mu_2} t+ \phi)\mathrm{e}^{-\lambda_2t}
  +A_{\mathrm{bg}}.
\end{multline}
Here $A_i$ and $\lambda_i$ ($i$ = $0, 1, 2$) represent the initial asymmetries and relaxation rates, respectively. 
$A_{\mathrm{bg}}$ was determined at a low temperature and fixed to $6.4$~\% both in the paramagnetic and ordered state. 
$\gamma_{\mu}$, $B_{\mu_1}$, and $B_{\mu_2}$ are the muon gyromagnetic ratio and the internal magnetic fields at the muon sites, respectively. 
$\phi$ is the phase of muon-spin precession determined by the transverse $\mu$SR measurement under $20$~G at $40$~K. 
The experimental data can be well reproduced by these fitting functions, as shown in Fig.~\ref{fig5}(a); 
this indicates that there are two major muon sites.

\begin{figure}[tbp]
  \begin{center}
  \includegraphics[width=7cm]{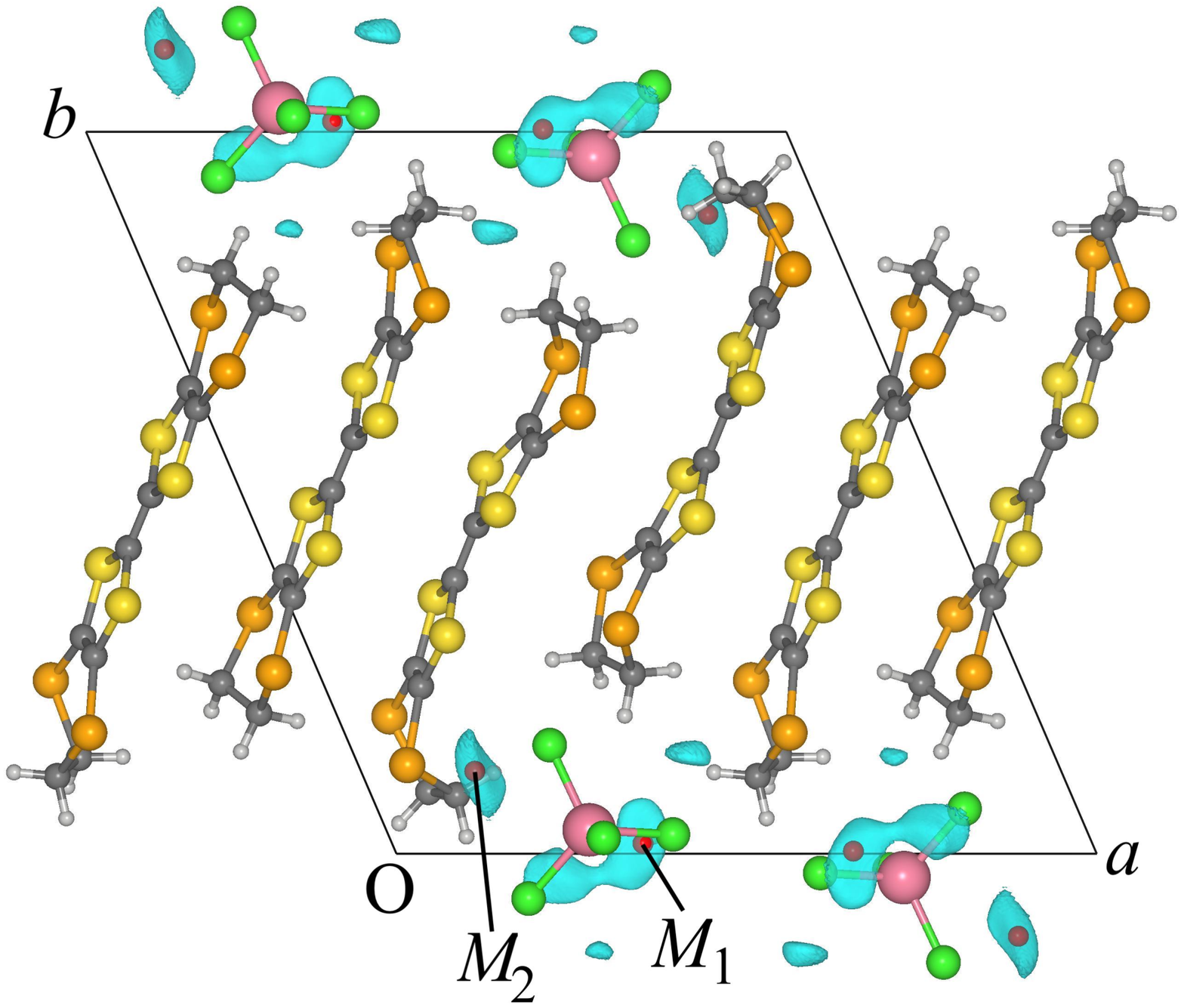}
  \end{center}
  \caption{
  Crystal structure of \lBEST, and electric minimum potential with isosurface of $37.8$~eV shown in the cyan region. 
  The most possible muon stopping sites are marked in red.
  }
  \label{muon}
  \end{figure}

The plot of $B_{\mu_2}$ against $B_{\mu_1}$ shown in Fig.~\ref{fig5}(b) indicates that they are proportional to each other. 
This result strongly suggests that the two observed rotational components correspond to the muons stopped at magnetically inequivalent sites and that the development of the magnetic moment is observed from different muon sites.
We discuss the positions of the two muon sites from the density functional theory (DFT) calculations performed within the Kohn--Sham approach using the projector augmented-waves formalism in the Vienna Ab-initio Simulation Packages (VASP) program \cite{Kresse1996,Kresse1996a}. 
The exchange--correlation function generalized gradient approximation, GGA-PW91, was used \cite{Perdew1996}. 
The ground-state charge densities were calculated by adopting the value of the crystal axis obtained in Sec.~\ref{crystalstructure}, and by using the $4\times4\times4$ $k$-point sampling, ultrasoft pseudopotentials, and plane-wave densities. 
The calculations were performed on the HOKUSAI supercomputer. 
Figure~\ref{muon} shows the crystal structure of \lBEST\ and the electric minimum potential with the isosurface of $37.8$~eV shown in cyan. 
Although there are several possible muon sites, we found two major sites: $M_1$ near GaCl$_4$, and $M_2$ near the ethylene groups. 
If the spin density is larger near the ethylene edge, it is likely that $M_1$ and $M_2$ are related to $B_{\mu_1}$ and $B_{\mu_2}$, respectively. 
Although \lBEST\ has a complex crystal structure, the major two muon sites observed in the present study can be consistently explained by the DFT calculations.

\begin{figure}[tbp]
\begin{center}
\includegraphics[width=8cm]{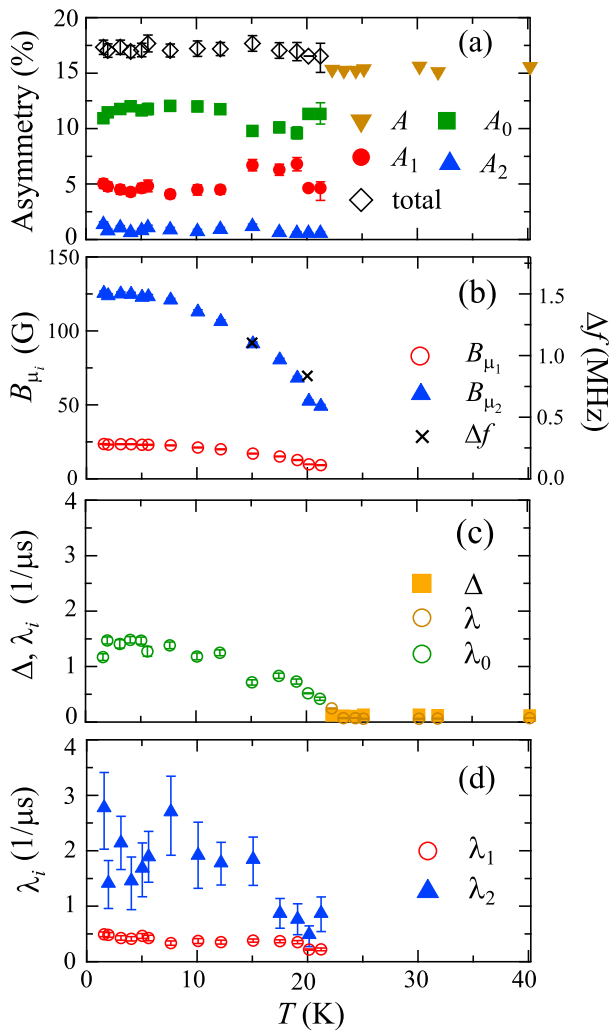}
\end{center}
\caption{
Temperature dependence of the parameters obtained by fitting the $\mu$SR time spectra: 
(a) Initial asymmetry, (b) internal fields at muon sites (left axis), (c) relaxation rates $\lambda$ and $\lambda_0$ and distribution width of the depolarization rate of the nuclear spin contribution $\Delta$, and (d) relaxation rates $\lambda_1$ and $\lambda_2$. 
Splitting widths of the NMR spectra $\Delta f$ in the AF state [defined in Fig.~\ref{fig7}(a)] are plotted in panel (b) with the right axis.
}
\label{fig6}
\end{figure}

Here, to obtain more accurate fitting results, we fitted the $\mu$SR time spectra again using the relationship $B_{\mu_2} = 5.338B_{\mu_1}$ obtained by the linear fitting of $B_{\mu_1}$ versus $B_{\mu_2}$.
Representative results of the fitting are shown in Fig.~\ref{fig5}(c), together with that of the paramagnetic phase.
The results of the fitting parameters are shown in Fig.~\ref{fig6}.
Figure~\ref{fig6}(a) shows the temperature dependence of the initial asymmetry.
The initial asymmetry is $15$~\% in the paramagnetic state, and it is distributed to each component by magnetic ordering; their total is $17$~\% in the AF state. 
Despite the different fitting functions below and above the magnetic ordering, the result of the almost unchanged total initial asymmetry verifies the validity of the analysis.

We discuss the volume fraction of long-range order based on the observed precession signals and total asymmetry below $T_{\rm N}$.
When a muon stops at a particular site in a long-range ordered state, 2/3 of the muon spins precess, and the remaining 1/3 do not because muons precess along one direction of the magnetic field. 
As shown in Fig.~\ref{fig6}(a), the precession components are $A_1 + A_2 \sim 6$~\%, and therefore, the fraction of muon spins undergoing internal magnetic fields in the long-range ordered state is $6$~\% $ \times~3/2 = 9$~\%. 
Since the total asymmetry is $17$~\%, the volume fraction of the magnetic ordered component is $9/17 \sim 53$~\%.
This is the lower limit estimated from the analysis, and the actual volume fraction is expected to be much larger because there are other minor muon sites shown in Fig.~\ref{muon}; $A_0$ is also considered to include components derived from long-range order even though they cannot be analyzed as precession signals.

The internal fields and relaxation rates increase with decreasing temperature [Fig.~\ref{fig6}(b) and (d)].
At sufficiently low temperatures below $T_{\rm N}$, the increase in the internal magnetic field is saturated, which is a characteristic of the change in the order parameter of the magnetic transition.
The extrapolated values of the two internal magnetic fields at $0$~K are $125$~G and $23.4$~G, respectively, and they are comparable to those of a typical organic antiferromagnet $\kappa$-(ET)$_2$Cu[N(CN)$_2$]Cl \cite{Ito2015}.
The relaxation rate $\lambda_0$ increases below $22$~K [Fig.~\ref{fig6}(c)] as well as the other rotational components. 
The value $\lambda_0$ = 1.5 $\mu$s$^{-1}$ estimated from Fig.~\ref{fig6}(c) at the lowest temperature is equivalent to the magnetic field $B$ = $17.6$~G, which is comparable to the value of $B_{\mu_1}$. 
This implies that the first term of Eq.~(\ref{eq2}) is from the same origin as the other rotational components. 

\subsection{NMR}
\label{NMR}
\begin{figure}[tbp]
\begin{center}
\includegraphics[width=8cm]{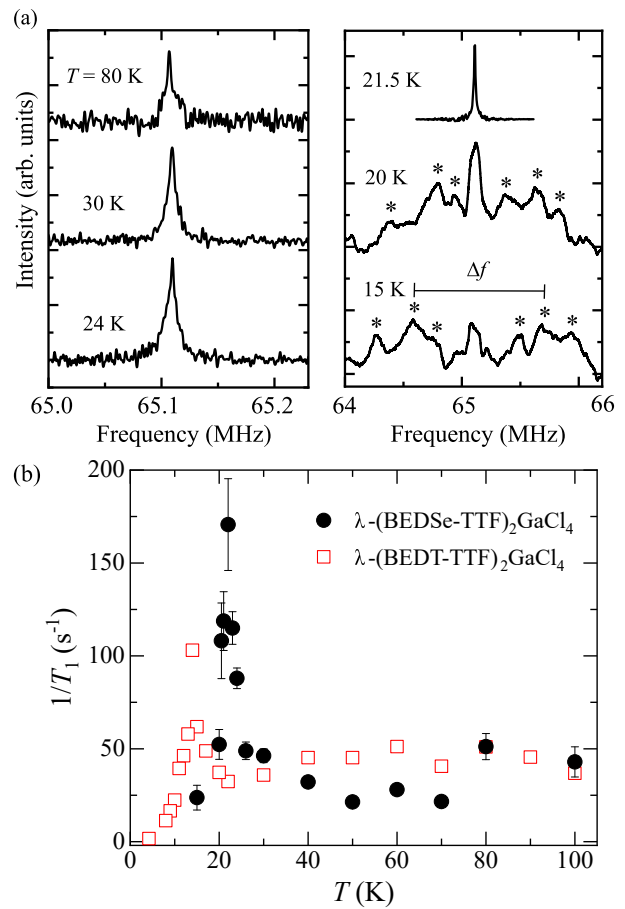}
\end{center}
\caption{
(a) $^{13}$C NMR spectra at several temperatures. 
The asterisks and $\Delta f$ indicate the splitting peaks and the width of the second peaks from outside, respectively. 
(b) Temperature dependence of $T_1^{-1}$ of \lBEST\ and \lET~\cite{Saito2018a}.
}
\label{fig7}
\end{figure}

In the NMR experiments, microscopic properties are probed similar to that in the $\mu$SR measurements. 
In addition, magnetic fluctuations can be detected from the $T_1$ measurement, which provides important insights into the nature of the magnetic state. 
The $^{13}$C-NMR method by $^{13}$C substitution of C=C atoms in the center of the TTF skeleton is a powerful method to investigate the electronic state, as has been established for the BEDT-TTF salts.
To conduct $^{13}$C-NMR measurements for \lBEST, we synthesized $^{13}$C enriched BEDSe-TTF molecules, as described in Sec.~\ref{exp}.
To the best of our knowledge, this is the first $^{13}$C-NMR measurement for BEDSe-TTF salts.

Figure~\ref{fig7}(a) shows the temperature evolution of the NMR spectra. 
In the paramagnetic state, a single peak was observed. 
As \lBEST\ has crystallographically independent BEDSe-TTF molecules I and II, each of which has two inequivalent $^{13}$C sites, four peaks are expected.
In the present experiment, a magnetic field was applied in the direction of the long axis of the BEDSe-TTF molecule, where the hyperfine coupling constant is small. 
As a result, the difference in the hyperfine coupling constants of each $^{13}$C site becomes small, resulting in a single overlapping spectrum.
Reflecting the presence of multiple $^{13}$C sites, the spectrum at $80$~K is not simply Lorentzian but shows a shoulder-like structure.
With decreasing temperature, the structure of the spectrum becomes less pronounced, but the general shape remains almost unchanged in the paramagnetic state above $24$~K.

The temperature dependence of $T_1^{-1}$ is shown in Fig.~\ref{fig7}(b). 
Because the spectrum consists of four peaks with slightly different hyperfine coupling constants, $T_1$ was determined by fitting the recovery of spin magnetization $M(t)$ using a stretched exponential function, $1-M(t)/M(\infty)$ = $\exp[-(t/T_1)^\beta]$, where $M(\infty)$ is the equilibrium spin magnetization at time $t \rightarrow \infty$ and $\beta$ is the stretched exponent. 
The recovery curves could be fitted by $\beta$ = $0.9$ for all temperatures. 
At high temperatures far above $T_{\rm N}$, $T_1^{-1}$ is constant as expected in a system with localized spins, indicating that the electronic state can be understood as a Mott insulator.
Below $26$~K, $T_1^{-1}$ drastically increases towards $T_{\rm N} \simeq 22$~K because of the critical slowing down, which evidences a second-order phase transition.
For comparison, the temperature dependence of $T_1^{-1}$ of \lET\ are also plotted in Fig.~\ref{fig7}(b) \cite{Saito2018a}.
The magnitude of $T_1^{-1}$ can be quantitatively compared because both experiments were performed under almost the same magnetic field direction and intensity and because the adjacent chemical environments around the central C=C atoms are the same between BEDSe-TTF and BEDT-TTF molecules.
The absolute values of $T_1^{-1}$ at high temperatures between both salts are comparable. 
As the values of $T_1^{-1}$ at high temperatures correlate with the magnitude of the exchange interaction, the present results indicate that exchange interactions of \lET\ and \lBEST\ are similar in order, although the lattice models may be different as discussed in Sec.~\ref{Spinsus}.

Below $20$~K, a drastic spectral splitting was observed, and this confirms the development of an internal magnetic field attributed to the magnetic ordering, which is consistent with the results of the $\mu$SR measurements.
The spectrum consists of a central peak and three symmetrically discrete spectra from the central peak.
Here, the discrete peaks are depicted by asterisks.
These results suggest that the spin structure is commensurate.
The commensurate spin structure with the central peak was similarly observed in \lET\ \cite{Saito2018a}, indicating that the AF spin structures between both salts are the same, although the different networks of $J$ are evaluated. 
The $T_1$ measurements in the AF state are performed on the central peak. 
$T_1^{-1}$ decreases steeply below $22$~K because of the decrease in the population of magnon excitations with decreasing temperature.

Analysis of the $\mu$SR spectra shows that the lower limit of the volume fraction of the long-range order is approximately half. 
However, the NMR measurements show that $1/T_1$ decreases steeply below $T_{\rm N}$ even in the central peak, suggesting that \lBEST\ exhibits an almost $100$~\% long-range AF order.
The $T_{\rm N}$ = $22$~K estimated by the NMR measurements is the same as the zero-field $T_{\rm N}$ estimated by the $\mu$SR measurements.
From the field-independent $T_{\rm N}$ and absence of weak ferromagnetic behavior as discussed in Sec.~\ref{Spinsus}, we suggested that no physical properties derived from the Dzyaloshinskii--Moriya interaction were observed, which would be expected when there is no inversion center between dimers as in $\kappa$-(ET)$_2$Cu[N(CN)$_2$]Cl \cite{Smith2004,Kagawa2008}.

By further decreasing the temperature to $15$~K, the spectral splitting is broadened.
Among the six symmetrically discrete peaks, the width between the most intense peaks was defined as $\Delta f$ [right side of Fig.~\ref{fig7}(a)].
To discuss the development of the internal magnetic field with decreasing temperature, their values at $15$ and $20$~K are shown in Fig.~\ref{fig6}(b). 
As the temperature dependence of the internal magnetic field observed at muon sites should be the same as that observed in NMR, the spectral splitting extrapolated to $0$~K is estimated to be $\sim 1.5$~MHz.
The splitting width of \lET\ is approximately $0.6$~MHz \cite{Saito2018a}, which is $2.5$ times smaller than that of \lBEST. 
Although the accurate ratio is difficult to obtain because of the broadness of the spectrum, the difference in the splitting width seems to correlate with the difference in $T_{\rm N}$. 

\subsection{Comparison of \lET~and \lBEST}
The present study demonstrated that \lBEST\ has almost the same value of $U/W$ as \lET. Furthermore, the behavior of the NMR spectrum and $T_1^{-1}$ is qualitatively the same between \lBEST\ and \lET, although the AF spin model inferred from the temperature dependence of $\chi_{\rm spin}$ is different.

In addition, there is a significant difference in $T_{\rm N}$.
$T_{\rm N}$ = $22$~K for \lBEST~is $1.7$ times larger than $T_{\rm N}$ = $13$~K for \lET. 
The $J$ of \lBEST\ and \lET~were estimated to be $J/k_B \sim 180$--$250$~K and $J/k_B \sim 98$~K \cite{Saito2018a} when the temperature dependence of $\chi_{\rm spin}$ is modeled by the 2D Heisenberg AF spin models.
These results simply suggest that $T_{\rm N}$ is approximately proportional to $J$. The large difference between $T_{\rm N}$ and $J/k_B$ in both salts suggest that a very weak interlayer interaction may suppress $T_{\rm N}$. In the case of quasi-2D Heisenberg antiferromagnets, the relationship between $T_{\rm N}$ and the intralayer and interlayer interactions have been discussed theoretically \cite{Yasuda2005}. The results suggest that the interlayer interaction of both \lET\ and \lBEST\ is far less than $J/1000$. Further, the literature also predicts that changes in $T_{\rm N}$ are sensitive to changes in intralayer interactions but insensitive to changes in interlayer interactions in cases where the interaction is extremely anisotropic. Therefore, even if the interlayer interactions were significantly different between the two salts within a sensible range, the twofold difference in $T_{\rm N}$ cannot be explained, and we conclude that it is because of the difference in the intralayer interactions. Another possibility is that a frustration effect may suppress the $T_{\rm N}$ in this series of the salts. Although this effect cannot be completely ruled out because \lBEST\ has a partially triangular lattice as described above, this effect would not explain the difference in $T_{\rm N}$ even if there were a frustration effect.

Here, we discuss the relative positions of \lBEST\ and \lET~in the universal phase diagram. 
$U/W$ is the primary parameter that should be considered; however, we cannot determine the relative positions from the $U/W$ values of \lET\ and \lBEST, between which there is no significant difference. 
The electrical resistivity measurements in $\kappa$-(BEDSe-TTF)$_2$Cu[N(CN)$_2$]Br indicate that the substitution of BEDT-TTF for BEDSe-TTF molecules has a negative pressure effect of approximately $0.15$~GPa \cite{Sakata1998}.
However, the analogy of the pressure effect from the results of $\kappa$-salt may be inappropriate because the molecular arrangement between the $\lambda$ and $\kappa$ phases is different.
Therefore, we discuss the position in the universal phase diagram from the change in $T_{\rm N}$ with pressure.
$^{13}$C-NMR experiments have been performed on \lET\ under pressure, which suggest that $T_{\rm N}$ decreases rapidly to $3$~K when a pressure of $0.4$~GPa is applied \cite{2021f}. Considering this result and the fact that \lET\ and \lBEST\ could be located in the same electronic phase, \lBEST, possessing a higher $T_{\rm N}$ than \lET, is located further to the negative pressure side. On the other hand, the electronic states of \lET\ under further pressure have not been detailed because of the existence of polymorphism and the difficulty of sample preparation. To establish the relationship between \lET\ and \lBEST\ in the universal phase diagram and reveal the electronic phases at higher pressures than AF state, we are currently conducting NMR experiments under pressure on \lBEST.

\begin{table}[tbp]
\caption{
Ground state and crystal preparation of $\lambda$-$D_2M$Cl$_4$ ($D$ = BETS, BEDT-STF, BEDT-TTF, BEDSe-TTF and $M$ = Ga, Fe). 
}
\begin{threeparttable}
\begin{ruledtabular}
\begin{tabular}{ccc}
& GaCl$_4^-$ & FeCl$_4^-$  \\ \hline
BETS & $\bigcirc$ SC & $\bigcirc$ AF (FISC) \\
BEDT-STF & $\bigcirc$ PI & ($\bigcirc$) AF \\
BEDT-TTF & $\triangle$ AF & $\times$ \\
BEDSe-TTF & $\doublecirc$ AF & $\doublecirc$ AF \\

\end{tabular}
\begin{tablenotes}
\item[] $\bigcirc$: mixture of the $\lambda$ and $\kappa$ phases, whereas $\kappa$-(STF)$_2$FeCl$_4$ has not been reported; $\doublecirc$: $\lambda$ phase only; 
$\triangle$: $\delta'$ and $\lambda$ phases ($\lambda$: minor product); $\times$: not reported
\end{tablenotes}
\end{ruledtabular}
\end{threeparttable}
\label{table4}
\end{table}

We mention the availability of various $\lambda$-$D_{2}M$Cl$_4$.
Table~\ref{table4} shows the combinations of donor molecule $D$ and anions $M$ = Ga and Fe along with the ground states of the compounds and the ease of obtaining $\lambda$-type salts (whether polymorphs are obtained simultaneously). 
$\lambda$-(BETS)$_2M$Cl$_4$ and $\lambda$-(STF)$_2M$Cl$_4$ are obtained together with the $\kappa$ phase \cite{Kobayashi1993a,Naito1997}, but they are easy to distinguish because of their different crystal shapes.
In contrast, \lET\ is difficult to distinguish from the $\delta$-type salt, and the $\lambda$ phase is a minor product \cite{Zorina2001,Zhang2013,Kurmoo1996}, which hampers the study of the AF phase using polycrystalline samples such as in $\mu$SR measurements. 
We found that \lBEST\ can be synthesized without other polymorphs, and this enabled the detailed measurements in the present study.
It has recently been demonstrated that the electronic system can be observed by polycrystalline $^{69,71}$Ga-NMR measurements on \lBETS\ \cite{Kobayashi2020b}; these experiments are easy to perform even under pressure, and therefore, \lBEST\ is also suitable for such experiments.

$\lambda$-(BETS)$_2$FeCl$_4$ exhibits a field-induced superconductivity \cite{Uji2001} and a strange metal--insulator transition with AF ordering \cite{Akiba2009a}, whereas $\lambda$-(STF)$_2$FeCl$_4$ shows a unique magnetic response, in which the magnetization processes between $\pi$ and $3d$ spin systems are different in the AF state \cite{Minamidate2018,Fukuoka2018,Fukuoka2020}.
In these Fe-containing systems, the $\pi$--$d$ interaction plays an essential role in the physical properties.
To understand the donor molecule substitution effect of the $\pi$--$d$ interaction, we are interested in the magnetic properties at the more negative pressure side, but the $\lambda$-(ET)$_2$FeCl$_4$ has not been reported.
$\lambda$-(BEDSe-TTF)$_2$FeCl$_4$, which is another possible salt located on the low pressure side, has been reported by Cui {\it et al.}, and magnetic susceptibility measurements suggest that it is paramagnetic down to $4$~K \cite{Cui2005a}. 
However, our research group has recently found that $\lambda$-(BEDSe-TTF)$_2$FeCl$_4$ exhibits an AF transition with a different magnetic process 
between $\pi$ and $3d$ spin systems \cite{Saito2022}. 
The result that \lBEST\ without $3d$ spins shows antiferromagnetism would be an important finding for the discussion of the $\pi$--$d$ interaction mechanism in $\lambda$-(BEDSe-TTF)$_2$FeCl$_4$.

\section{Summary}
We investigated the structural and magnetic properties of \lBEST\ to determine whether this material can be placed on the universal phase diagram of $\lambda$-$D_2$GaCl$_4$. 
The systematic band calculations for $D$ = BEDSe-TTF, BEDT-TTF, and BETS salts suggested that \lBEST\ is a Mott insulator as well as \lET\ and that there is a Mott transition between them and \lBETS. 
Further, we found that the network of $J$ in the \lBEST\ salt consists of a combination of triangular and square ladders.
The broad peaks observed in the temperature dependence of $\chi_{\rm spin}$ are intermediate between those in the triangular and square lattice Heisenberg AF spin models, which is consistent with the network of $J$.
From a microscopic viewpoint, the development of the internal magnetic fields was observed from muon precession signals, which is definitive evidence that \lBEST\ exhibits AF ordering. 
In the $^{13}$C-NMR measurement, we observed the divergent behavior of $(T_1T)^{-1}$ towards $22$~K, below which the NMR spectra split discretely, retaining the central peak, as observed in \lET. 
These behaviors are qualitatively the same as those of \lET, which sugges that both salts are in the same electronic phase in the universal phase diagram of $\lambda$-type salts.
Further, \lBEST\ can be synthesized without polymorphism, unlike \lET.
These features and the present results promote the understanding of the nature of the electronic states located at lower pressure than the SC phase in the phase diagram, e.g., whether the nonmagnetic ordered phase between the AF and SC phases is intrinsic by the experiments under pressure.

\begin{acknowledgments}
The authors would like to thank Prof. R. Kato for his advice on the synthesis of the BEDSe-TTF molecule. This work was partly supported by Hokkaido University, Global Facility Center (GFC), Advanced Physical Property Open Unit (APPOU), funded by MEXT under ``Support Program for Implementation of New Equipment Sharing System'' (JPMXS0420100318).
This work was also partially supported by the Japan Society for the Promotion of Science KAKENHI Grant Numbers 20K14401, 19K03758, and 21K03438.
\end{acknowledgments}


\begin{thebibliography}{75}
\bibitem [{Pha()}]{Phasediagram}%
    \BibitemOpen
    \href@noop {} {}\bibinfo {note} {The relative positions between $\lambda$-ET
    and $\lambda$-STF and between $\lambda$-STF and $\lambda$-BETS were
    determined from the pressure variation of the temperature dependence of
    electrical resistivity [24] and the pressure dependence of superconducting
    transition temperature $T_c$ [21,27], respectively.}\BibitemShut {Stop}%
\bibitem [{Huc()}]{Huckelparameter}%
    \BibitemOpen
    \href@noop {} {}\bibinfo {note} {The $\xi$ exponent (ionization potential
    (eV)) for atomic orbitals are as follows. Se : 4d 2.44 (-20.0), 4p 2.07
    (-10.8), 4s 1.5 (-5.44). S : 3d 2.122 (-20.0), 3p 1.827 (-11.0), 3s 1.5
    (-5.44). C : 2p 1.625 (-21.4), 2s 1.625 (-11.4). H : 1s 1.3
    (-13.6).}\BibitemShut {Stop}%
\end{thebibliography}
%
\end{document}